\documentclass[
reprint,
superscriptaddress,
amsmath,amssymb,
aps,
]{revtex4-2}
\usepackage{booktabs}
\usepackage{array}
\usepackage{graphicx} 
\usepackage{subfigure}
\usepackage{dcolumn}
\usepackage{bm}
\usepackage[colorlinks=true, linkcolor=blue, citecolor=blue]{hyperref}
\usepackage{cleveref}
\usepackage{xcolor}
\usepackage{float}
\usepackage{titlesec}
\usepackage{tabularx}
\titlespacing*{\section}{0pt}{12pt}{6pt}
\titlespacing*{\subsection}{0pt}{6pt}{4pt}

\hypersetup{colorlinks=true,linkcolor=blue,filecolor=blue,urlcolor=blue,citecolor=blue}
\newcommand{\doubletoprule}{%
	\hline
	\noalign{\vspace{1pt}}
	\hline
}

\newcommand{\doublebottomrule}{%
	\hline
	\noalign{\vspace{1pt}}
	\hline
}
\begin{document}

\preprint{APS/123-QED}

\title{Semi-Device-Independent Quantum Random Number Generator Resistant to General Attacks }

\author{Zhenguo Lu}
\affiliation{State Key Laboratory of Quantum Optics Technologies and Devices, Institute of Opto-Electronics, Shanxi University, Taiyuan 030006, China
}
\affiliation{Collaborative Innovation Center of Extreme Optics, Shanxi University, Taiyuan 030006, China
}
\author{Jundong Wu}
\affiliation{State Key Laboratory of Quantum Optics Technologies and Devices, Institute of Opto-Electronics, Shanxi University, Taiyuan 030006, China
}
\affiliation{Collaborative Innovation Center of Extreme Optics, Shanxi University, Taiyuan 030006, China
}

\author{Yu Zhang}
\affiliation{State Key Laboratory of Quantum Optics Technologies and Devices, Institute of Opto-Electronics, Shanxi University, Taiyuan 030006, China
}
\affiliation{Collaborative Innovation Center of Extreme Optics, Shanxi University, Taiyuan 030006, China
}
\author{Shaobo Ren}
\affiliation{State Key Laboratory of Quantum Optics Technologies and Devices, Institute of Opto-Electronics, Shanxi University, Taiyuan 030006, China
}
\affiliation{Collaborative Innovation Center of Extreme Optics, Shanxi University, Taiyuan 030006, China
}
\author{Xuyang Wang}
\affiliation{State Key Laboratory of Quantum Optics Technologies and Devices, Institute of Opto-Electronics, Shanxi University, Taiyuan 030006, China
}
\affiliation{Collaborative Innovation Center of Extreme Optics, Shanxi University, Taiyuan 030006, China
}
\affiliation{Hefei National Laboratory, Hefei 230088, China
}
\author{Hongyi Zhou}
\affiliation{State Key Laboratory of Processors, Institute of Computing Technology, Chinese Academy of Sciences, 100190, Beijing, China. 
}
\author{Yongmin Li}
\email[]{yongmin@sxu.edu.cn}
\affiliation{State Key Laboratory of Quantum Optics Technologies and Devices, Institute of Opto-Electronics, Shanxi University, Taiyuan 030006, China
}
\affiliation{Collaborative Innovation Center of Extreme Optics, Shanxi University, Taiyuan 030006, China
}
\affiliation{Hefei National Laboratory, Hefei 230088, China
}

\hypersetup{
	colorlinks=true,
	linkcolor=blue,
	filecolor=blue,
	urlcolor=blue,
	citecolor=blue,
}

\begin{abstract}
 Quantum random number generators (QRNGs) produce true random numbers based on the inherent randomness of quantum theory, rendering them a foundational segment of quantum cryptography. Distinguished from trusted-device QRNGs whose security depends on characterized devices, semi-device-independent (semi-DI) QRNGs permit partial devices to be defective or even maliciously manipulated, which achieves a good trade-off between generation rate and security. In this paper, we propose a semi-DI QRNG that resists general attacks while accounting for finite-size effects. The protocol requires no rigorous characterization of the source and measurement devices other than limiting the energy of the emitted states, significantly reducing the demands on practical QRNG systems. Leveraging the tight Kato’s inequality for correlated variables, we show that our protocol generates more randomness than it consumes. Furthermore, we demonstrate the scheme on a continuous-variable system with ternary inputs of states. Heterodyne detection is employed to enable phase compensation through data  postprocessing, alleviating the stringent requirement on system stability. The system operates at 100 MHz, achieving a net random number generation rate of 1.165 Mbps at $5.3\times10^9$ rounds. Our work offers a promising approach to achieve both the robust security and  high generation rate with a simple experimental setup.
\end{abstract}

\maketitle

\section{\label{sec:level1}INTRODUCTION}
\hspace*{1em}True random numbers are essential for cryptography, simulations and machine learning. The primary metrics of random numbers are statistical uniformity and unpredictability. The random number generation based on classical physical processes is inherently predictable, whereas quantum random number generators (QRNGs) leverage the probabilistic nature of quantum state collapse to produce true random numbers that show uniformity and unpredictability  \cite{herrero-collantesQuantumRandomNumber2017a}.

QRNGs with trusted  apparatus are the most common approach to generate random numbers and known  for their high generation speed and easy implementation \cite{shenPracticalQuantumRandom2010,hawMaximizationExtractableRandomness2015a,Wayne20022009,qiHighspeedQuantumRandom2010,symulRealTimeDemonstration2011,gehringHomodynebasedQuantumRandom2021a,bai188GbpsRealtime2021}. In practical implementation of QRNGs, side-channel information can be leaked to potential adversaries due to the imperfect  devices, which threatens the security and privacy of random generated numbers.

Device-independent protocols have been  proven secure without any assumptions about the devices used \cite{pironioRandomNumbersCertified2010,liuDeviceindependentQuantumRandomnumber2018a,shalmDeviceindependentRandomnessExpansion2021,brownFrameworkQuantumSecureDeviceIndependent2020,DI-RA-PR,FullRandomnessfromArbitrarilyDeterministicEvents,Realisticnoise-tolerantrandomnessamplificationusinginitenumbeofdevices,colbeckFreeRandomnessCan2012,kulikov2024deviceindependentrandomnessamplification}. Randomness expansion and randomness amplification are two families of DI protocols. The former requires a perfect randomness input to select measurement settings in a Bell test \cite{pironioRandomNumbersCertified2010,liuDeviceindependentQuantumRandomnumber2018a,shalmDeviceindependentRandomnessExpansion2021,brownFrameworkQuantumSecureDeviceIndependent2020}, whereas the latter can tolerate an imperfect randomness input \cite{DI-RA-PR,FullRandomnessfromArbitrarilyDeterministicEvents,Realisticnoise-tolerantrandomnessamplificationusinginitenumbeofdevices,colbeckFreeRandomnessCan2012,kulikov2024deviceindependentrandomnessamplification}. However, DI protocols require the loophole-free violation of a Bell inequality, which are experimentally challenging to implement and suffer from low random number generation rates.  To balance security and practicality, semi-DI QRNGs that partially trust the device have been developed \cite{linCertifiedRandomnessUntrusted2022,pivoluskaSemideviceindependentRandomNumber2021,caoSourceIndependentQuantumRandom2016a,marangonSourceDeviceIndependentUltrafastQuantum2017b,avesaniSourcedeviceindependentHeterodynebasedQuantum2018a,drahiCertifiedQuantumRandom2020,avesaniUnboundedRandomnessUncharacterized2022a,liOnchipSourcedeviceindependentQuantum2024,duSourceindependentQuantumRandom2025,zhouContinuousVariableSourceIndependentQuantum2025,chaturvediMeasurementdeviceindependentRandomnessLocal2015a,caoLosstolerantMeasurementdeviceindependentQuantum2015a,nieExperimentalMeasurementdeviceindependentQuantum2016a,PhysRevA.95.062305,wangProvablysecureQuantumRandomness2023a}. Currently, there are two main types of semi-DI QRNGs: source-independent (SI) \cite{caoSourceIndependentQuantumRandom2016a,marangonSourceDeviceIndependentUltrafastQuantum2017b,avesaniSourcedeviceindependentHeterodynebasedQuantum2018a,drahiCertifiedQuantumRandom2020,avesaniUnboundedRandomnessUncharacterized2022a,liOnchipSourcedeviceindependentQuantum2024,duSourceindependentQuantumRandom2025,zhouContinuousVariableSourceIndependentQuantum2025} and measurement-device-independent (MDI) QRNGs \cite{chaturvediMeasurementdeviceindependentRandomnessLocal2015a,caoLosstolerantMeasurementdeviceindependentQuantum2015a,nieExperimentalMeasurementdeviceindependentQuantum2016a,PhysRevA.95.062305,wangProvablysecureQuantumRandomness2023a}, which avoid the precise characterization of the source and measurement devices, respectively.

Recently, a variety of SI (MDI) QRNGs protocols with relaxing security assumptions on the  measurement (source) devices have been proposed \cite{linSecurityAnalysisImprovement2020,liImpactImperfectSampling2024,nieMeasurementdeviceindependentQuantumRandom2024a,wuMeasurementDeviceIndependentQuantumRandomNumber2023a,braskMegahertzRateSemiDeviceIndependentQuantum2017b,ruscaFastSelftestingQuantum2020a,tebyanianSemideviceIndependentRandomness2021a,avesaniSemiDeviceIndependentHeterodyneBasedQuantum2021a,himbeeckSemideviceindependentFrameworkBased2017}. For example, security risks are examined in SI-QRNG arising from afterpulse effects and detection efficiency mismatches in single-photon detectors \cite{linSecurityAnalysisImprovement2020}. Randomness lower bounds in MDI QRNGs with defective single-photon sources have been derived using the decoy state method \cite{nieMeasurementdeviceindependentQuantumRandom2024a,wuMeasurementDeviceIndependentQuantumRandomNumber2023a}. Semi-DI QRNGs that constrain the overlap or energy of prepared quantum states have also been investigated \cite{braskMegahertzRateSemiDeviceIndependentQuantum2017b,ruscaFastSelftestingQuantum2020a,tebyanianSemideviceIndependentRandomness2021a,avesaniSemiDeviceIndependentHeterodyneBasedQuantum2021a,himbeeckSemideviceindependentFrameworkBased2017}. These studies have relaxed the assumptions of semi-DI frameworks and improve the security while broadening their applicability. However, the independent and identically distributed (i.i.d.) assumption is stringently imposed on the measurement devices  in current MDI QRNGs. An adversary can employ correlated measurement strategies in different random number generation rounds at the measurement side, which are known as general attacks \cite{caoLosstolerantMeasurementdeviceindependentQuantum2015a}. This will result in potentially correlated outputs and thus enhancing the adversary’s ability to guess the random sequences. A recently proposed numerical framework for security analysis provides lower bounds on the secure randomness of semi-DI QRNGs under general attacks \cite{zhouNumericalFrameworkSemideviceindependent2023a}.

In this paper, we present a multi-input semi-DI QRNG based on heterodyne detection and demonstrate its security under general attacks. The measurement apparatus is treated as a black box that hides performed quantum measurements to the user. Crucially, our protocol allows an adversary to perform correlated quantum operations at the measurement side without being constrained by the i.i.d. assumption, and takes into account finite-size effects. The upper bound on the energy of the prepared quantum states of the source is constrained without requiring full knowledge of the states emitted in each round. We experimentally validate the semi-DI QRNG with three inputs and multiple outputs in continuous-variable (CV) system with components that are commercially off-the-shelf. Furthermore, as heterodyne detection realizes full phase-space tomography, the phase compensation can be enabled by data postprocessing and a real-time phase stabilization system is unnecessary, which reduces the system complexity and enhances the practicability of the protocol. Finally, since our protocol is, in essence, a form of quantum randomness expansion, we demonstrate that it is capable of generating more randomness than it consumes.

\section{THE PROTOCOL}
\subsection{\label{sec:level2}Semi-DI QRNG with ternary inputs}
\hspace*{1em}Our protocol consists of three steps: (1)  preparations and measurements of  quantum states, (2) estimation of the genuinely quantum entropy, (3) randomness extraction. In step (1),  a user randomly designates a prepare-and-measure round as either a test round with probability $p_t$ or a generation round with probability $1-p_t$. In a test round, the user randomly prepares quantum states $\rho_x$ ($x \in \{1,2,3\}$) with probability $p_x$, which are transmitted to an untrusted measurement station. The measurement device  measures the quantum states and outputs the results $y \in \{1,2,3,\cdots,d\}$. In a generation round, the user prepares a fixed state  $\rho_3$, which is also measured by the measurement device to yield the outcomes $y \in \{1,2,3,\cdots,d\}$. The above process is repeated  $N$ times, and we record the count $N_{y|x}$ representing the number of measurement outcome $y$ given input state $\rho_x$ during test rounds. In step (2), the user calculates the min-entropy according to Eq.~\eqref{eq:min_entropy_fin} for finite-size case or Eq.~\eqref{eq:min_entropy} for asymptotic limit. In step (3),  based on evaluated min-entropy, secure random numbers can be extracted~\cite{tomamichelLeftoverHashingQuantum2011} by postprocessing the data from the generation rounds, where a strong extractor is applied.

Similar to previous work~\cite{avesaniSemiDeviceIndependentHeterodyneBasedQuantum2021a}, our protocol does not require the exact characterization of the source, instead, it constrains the lower bound of the overlap of quantum states across different rounds by limiting the energy of the prepared states as follows:
\begin{equation}
|\langle \psi_x | \psi_{x'} \rangle| \geq 1 - 2\mu, \quad x, x' \in \{1,2,3\}
\label{eq:state_overlap}
\end{equation}
where $\mu$ is the upper bound of the mean photon number of the emitted states that can be considered pure states. 

In our protocol, (1) we assume an i.i.d. source and suppose that neither the source nor the measurement shows a quantum correlation with the environment, while a classical correlation can exist. (2) Following the research of semi-DI QRNGs \cite{braskMegahertzRateSemiDeviceIndependentQuantum2017b,tebyanianSemideviceIndependentRandomness2021a,avesaniSemiDeviceIndependentHeterodyneBasedQuantum2021a}, we emphasize that the generation of input random numbers (used to select the state and designate the test or generation round in our case) must be independent of both the devices and any adversary.

We now briefly review min-entropy evaluation under collective attacks in the asymptotic limit. In this scenario, the adversary Eve executes different but uncorrelated quantum measurements in each round  \cite{caoLosstolerantMeasurementdeviceindependentQuantum2015a}. Under the protocol’s assumptions, Eve can manipulate the source and the measurement apparatus to disrupt output randomness. Let's denote Eve’s attack strategy by $s$, then the untrusted measurement apparatus can be characterized by a set of  $d$-dimensional positive operator-valued measure (POVM) operators $\{ M^s_y \}$, and the transmitted quantum states can be denoted as $|\psi^s_x\rangle$. The lower bound of the inner product of states in Eq.~(\ref{eq:state_overlap}) is taken as the worst estimate of randomness destruction. Hence, the states $|\psi^s_x\rangle$ can always be expanded in terms of a set of basis vectors $\{ |1\rangle, |2\rangle, |3\rangle \}$ in the Hilbert space as follows ~\cite{tebyanianSemideviceIndependentRandomness2021a}:

\begin{equation}
\centering
\begin{aligned}
| \psi_1 \rangle &= | 1 \rangle, \\
| \psi_2 \rangle &= \beta | 1 \rangle + \sqrt{1 - \beta^2} | 2 \rangle, \\
| \psi_3 \rangle &= \beta | 1 \rangle + \beta \sqrt{\frac{1 - \beta}{1 + \beta}} | 2 \rangle + \sqrt{\frac{1 + \beta - 2\beta^2}{1 + \beta}} | 3 \rangle,
\end{aligned}
\end{equation}
where $\beta = 1 - 2\mu$. Then, Eve's guessing probability $P_g$ can be expressed as
\begin{equation}
P_g = \max_{\{ M^s_y \}} \sum_{s}  p(s) \max_y \mathrm{tr}(\rho_3 M^s_y),
\label{eq:P_g}
\end{equation}
where $p(s)$ represents the probability of Eve adopting measurement strategy $s$. As shown in Eq.~(\ref{eq:P_g}), Eve maximizes her guessing power by optimizing the measurement $M^s_y$ and the corresponding distribution $p(s)$. However, as the number of strategies $s$ can be infinite, Eq.~(\ref{eq:P_g}) is difficult to compute. Following the approach  in Ref.~\cite{bancalMoreRandomnessSame2014a,braskMegahertzRateSemiDeviceIndependentQuantum2017b}, we divide the measurement operators $M^s_y$ into $d$ groups, each with elements $M^{\tilde{s}}_y$ ($\tilde{s} \in \{1,2,3,\cdots,d\}$) and satisfying $\max_y \mathrm{tr}(\rho_3 M^{\tilde{s}}_y) = \mathrm{tr}(\rho_3 M^{\tilde{s}}_{\tilde{s}})$. Therefore, Eq.~(\ref{eq:P_g}) can  be rewritten as follows:
\begin{equation}
P_g = \max_{\{\Pi_y^{\tilde{s}}\}} \sum_{\tilde{s}} \mathrm{tr}(\rho_3 \Pi^{\tilde{s}}_{\tilde{s}}),
\label{eq:P_g_rewritten}
\end{equation}
where $\Pi_y^{\tilde{s}} = p(\tilde{s}) M^{\tilde{s}}_y$. Meanwhile, the operators $\Pi_y^{\tilde{s}}$ are subject to the following constraints:
\begin{equation}
\begin{aligned}
&\Pi_y^{\tilde{s}} \succeq 0, \\
&(\Pi_y^{\tilde{s}})^\dagger = \Pi_y^{\tilde{s}}, \\
&\sum_y \Pi_y^{\tilde{s}} = \frac{1}{3} \mathrm{tr}\left(\sum_y \Pi_y^{\tilde{s}}\right)I, \\
&\sum_{\tilde{s}=1}^d \mathrm{tr}(\rho_x \Pi^{\tilde{s}}_y) = p(y|x).
\end{aligned}
\label{eq:constraints}
\end{equation}
Here, $p(y|x) = \lim\limits_{N \to +\infty} \frac{N_{y|x}}{p_tp_x N}$ is the experimentally observed conditional probability. Combining Eqs.~(\ref{eq:P_g_rewritten}) and (\ref{eq:constraints}), the guessing probability $P_g$ can be determined by solving a semidefinite programming (SDP). The min-entropy describing the lower bound of secure randomness is then calculated as follows:
\begin{equation}
H_{\min} = -\log_2 P_g.
\label{eq:min_entropy}
\end{equation}
Note that in the asymptotic limit ($N \to +\infty$), the statistical fluctuations in $p(y|x)$ are negligible and then the optimal probability of test rounds $p_t$ tends to zero.  Hence, randomness consumption of the protocol vanishes.
\subsection{Security analysis against general attacks
}
\label{sec:theory}
\hspace*{1em}As previously mentioned, the adversary Eve randomly selects a measurement strategy in each round to obtain the side information of the QRNG system. However, by performing correlated quantum operations across multiple rounds, Eve can insert predefined correlations in the outputs, enhancing her guessing ability. Equivalently, Eve executes correlated POVM measurements on the quantum states in different rounds~\cite{caoLosstolerantMeasurementdeviceindependentQuantum2015a}. Moreover, due to the limited number of rounds, the estimation of conditional probabilities $p(y|x)$ is inaccurate, making the system vulnerable. Fortunately, analogous to the handling of statistical fluctuations in the source-trusted case \cite{zhouNumericalFrameworkSemideviceindependent2023a}, we can introduce a stochastic process that witnesses Eve’s correlated operations. By employing concentration inequalities to account for correlated variables, the loss of private randomness can then be quantified in both the scenarios.

As described in Ref.~\cite{zhouNumericalFrameworkSemideviceindependent2023a}, we first solve the dual problem~\cite{boydConvexOptimization2004} of Eq.~(\ref{eq:P_g_rewritten}):
\begin{equation}
\begin{aligned}
& \min_{\{H^{\tilde{s}},\omega_{xy}\}} - \sum_{x,y} \omega_{xy} p'(y|x) \\
\text{s.t.} \quad & (H^{\tilde{s}})^\dagger = H^{\tilde{s}}, \\
& \sum_x \rho_x \left(\delta_{3,x} \delta_{\tilde{s},y} + \omega_{xy}\right) + H^{\tilde{s}} \\
& \qquad - \frac{1}{3} \mathrm{tr}(H^{\tilde{s}})I \preceq 0, \\
\end{aligned}
\label{eq:dual_problem}
\end{equation}

to obtain a set of solutions $\{\omega_{xy}'\}$ that satisfy
\begin{equation}
\begin{aligned}
& \sum_x \rho_x \left(\delta_{3,x} \delta_{\tilde{s},y} + \omega_{xy}'\right) + H^{\tilde{s}} \\
& \qquad - \frac{1}{3} \mathrm{tr}(H^{\tilde{s}})I \preceq 0, \\
\end{aligned}
\label{eq:solution_set}
\end{equation}

As the conditional probabilities $p'(y|x)$ substituted into Eq.~(\ref{eq:dual_problem}) are theoretical values computed using a physical model, they are independent of Eve’s attack behavior. $p'(y|x)$ can also be regarded as nominal values ~\cite{zhouNumericalMethodFinitesize2022a}.

We now define a stochastic process $\{W_n\}$ as follows:
\vspace{-\baselineskip} 
\begin{flushleft}  
\begin{equation}
\begin{aligned}
&W_n = \\  
&\left\{
\begin{array}{@{}l@{\quad}l@{}}
\dfrac{1}{1 - p_t}, & 
    \begin{aligned}
    &\text{if the } n\text{-th round is a generation round and} \\
    &\text{Eve successfully guesses the output}
    \end{aligned} \\[3ex]
\dfrac{\omega_{xy}'}{p_t p_x}, & 
    \begin{aligned}
    &\text{if the } n\text{-th round is a test round and} \\
    &\text{the measured output is } y \text{ when emitting } \rho_x
    \end{aligned} \\[3ex]
0, & \text{otherwise}
\end{array}
\right.
\end{aligned}
\label{eq:Wn_definition}
\end{equation}
\end{flushleft}
Let $\{\mathcal{F}_{n-1}\}$ be a filtration containing information regarding $W_n$ from the first round to the $(n-1)$-th round, we then have
\begin{equation}
\begin{split}
E(W_n|\mathcal{F}_{n-1}) &= \sum_{\tilde{s}}  \mathrm{tr}(\rho_3 \Pi_{\tilde{s}}^{\tilde{s},\mathcal{F}_{n-1}}) \\
&+ \sum_{x,y,\tilde{s}} \omega_{xy}' \mathrm{tr}(\rho_x \Pi_{y}^{\tilde{s},\mathcal{F}_{n-1}}),
\end{split}
\label{eq:conditional_expectation}
\end{equation}
where $\Pi_{y}^{\tilde{s},\mathcal{F}_{n-1}}$ is the Eve’s quantum operation correlated with the first $n-1$ rounds that meets the following normalization condition:
\begin{equation}
\sum_y \Pi_{y}^{\tilde{s},\mathcal{F}_{n-1}} = \frac{1}{3} \mathrm{tr}\left(\sum_y \Pi_{y}^{\tilde{s},\mathcal{F}_{n-1}}\right)I.
\label{eq:normalization}
\end{equation}
Combining Eqs.~\eqref{eq:solution_set}, \eqref{eq:conditional_expectation} and \eqref{eq:normalization}, we obtain
\begin{equation}
E(W_n|\mathcal{F}_{n-1}) \leq 0.
\label{eq:expectation_bound}
\end{equation}

By the definition of $W_n$, we have
\begin{equation}
\sum_{n=1}^N W_n = \frac{N_g}{1 - p_t} + \sum_{x=1}^3 \sum_{y=1}^d \frac{\omega_{xy}'}{p_x p_t} N_{y|x},
\label{eq:Wn_sum}
\end{equation}
where $N_g$ is the number of successful guesses.

Herein, to achieve a tighter bound of randomness, we employs  the Kato’s inequality derived in the Ref.~\cite{zhouNumericalMethodFinitesize2022a} instead of the Azuma’s inequality used in Ref.~\cite{zhouNumericalFrameworkSemideviceindependent2023a}. As the first step in applying Kato’s inequality, $W_n$ needs to be normalized into the range [0,1], namely
\begin{equation}
W_n^{\text{nor}} = \frac{W_n - W_{\min}}{W_{\max} - W_{\min}},
\label{eq:normalize}
\end{equation}
where $W_{\min} = \min W_n$ and $W_{\max} = \max W_n$. By applying Kato's inequality \cite{kato2020concentrationinequalityusingunconfirmed} to $W_n^{\text{nor}}$, we have
\begin{equation}
\begin{split}
\sum_{n=1}^{N} W_n^{\text{nor}} - \sum_{n=1}^{N} E(W_n^{\text{nor}} \mid \mathcal{F}_{n-1}) \\
\leq \left[ b + a \left( 2\frac{\sum_{n=1}^{N} W_n^{\text{nor}}}{N} - 1 \right) \right] \sqrt{N},
\end{split}
\label{eq:kato-in}
\end{equation}
which holds with a probability of at least $1-\epsilon$. Here, the parameters $a \in \mathbb{R}$ and $b\geq 0$ satisfy 
\begin{equation}
\exp\left(-\frac{2(b^2 - a^2)}{\left(1 - \frac{4a}{3\sqrt{N}}\right)^2}\right) = \epsilon.
\label{eq:a-b-eps}
\end{equation}

From Eq.~\eqref{eq:expectation_bound}, we have
\begin{equation}
\begin{split}
\sum_{n=1}^{N} E(W_n^{\mathrm{nor}} \mid \mathcal{F}_{n-1}) 
    &= \sum_{n=1}^{N} \frac{E(W_n - W_{\min} \mid \mathcal{F}_{n-1})}{W_{\max} - W_{\min}} \\
    &\leq -\frac{NW_{\min}}{W_{\max} - W_{\min}} .
\end{split}
\label{eq:E-NEW}
\end{equation}
Thus, given $\sqrt{N} - 2a > 0$, Eq.~\eqref{eq:kato-in} can be rewritten as
\begin{equation}
W_{s}^{\mathrm{nor}} \leq \frac{\sqrt{N}}{\sqrt{N} - 2a} 
\left[ (b - a)\sqrt{N} - \frac{N W_{\min}}{W_{\max} - W_{\min}} \right],
\label{eq:ineq-rewrite}
\end{equation}
where $W_s^{\mathrm{nor}} = \sum_{n=1}^N W_n^{\mathrm{nor}}$. Combining Eqs.~\eqref{eq:Wn_sum}, \eqref{eq:normalize} and \eqref{eq:ineq-rewrite}, we obtain
\begin{equation}
N_g \leq \left[ \Delta\bigl(N, \epsilon(a, b)\bigr) 
- \sum_{x=1}^3 \sum_{y=1}^d \frac{\omega'_{xy}}{p_x p_t} N_{y|x} \right] (1-p_t),
\label{eq:Ng}
\end{equation}
where 
\begin{equation}
\Delta \! \left( N, \epsilon(a, b) \right) 
= \frac{N \bigl[ (b - a) W_{\max} - (b + a) W_{\min} \bigr]}{\sqrt{N} - 2a}.
\label{eq:delta-f}
\end{equation}
Therefore, the min-entropy $H'_{\min}$ per generation round against general attacks is determined as
\begin{equation}
H'_{\min} = -\log_2 \left(\frac{(N_g)_{\max}}{N(1 - p_t)}\right).
\label{eq:min_entropy_fin}
\end{equation}
Taking into account the randomness consumed in the protocol, the net randomness generation rate $R_{\mathrm{net}}$ of our protocol is defined as
\begin{equation}
R_{\mathrm{net}} = R_{\mathrm{gross}} - R_{\mathrm{in}},
\label{eq:net-rate}
\end{equation}
where $R_{\mathrm{gross}}=(1-p_t)H'_{\min}$ is the gross randomness  produced by the protocol per round, and $R_{\mathrm{in}}$ is the input randomness per round. Details on the evaluation of $R_{\mathrm{in}}$ are provided in Sec.~\ref{sec:citeref}.
\begin{figure}[t]
	\centering
	\subfigure[]{
		\includegraphics[width=0.45\columnwidth]{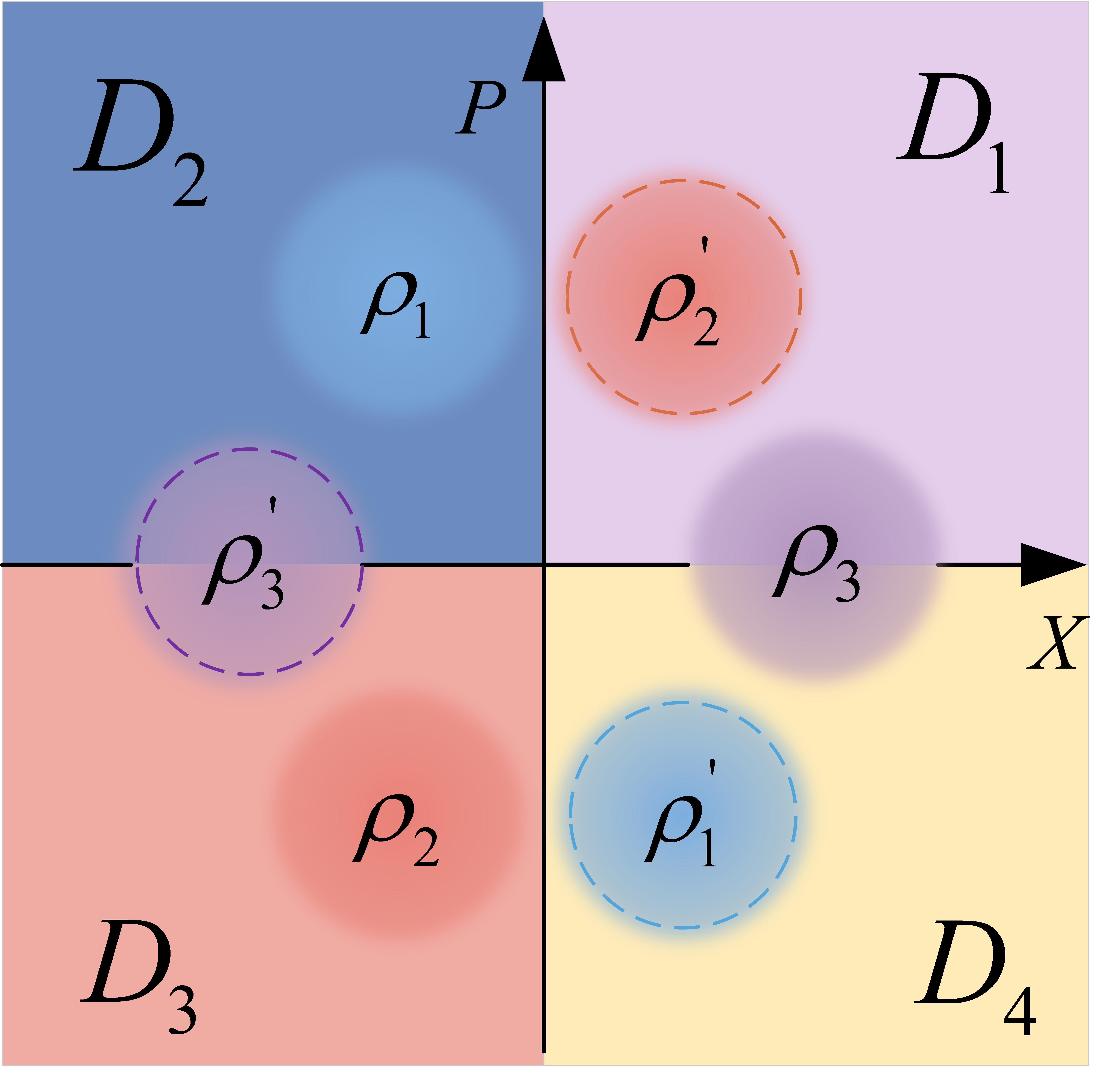}
		
		\label{fig:1a}
	}
	\hfill
	\subfigure[]{
		\includegraphics[width=0.46\columnwidth]{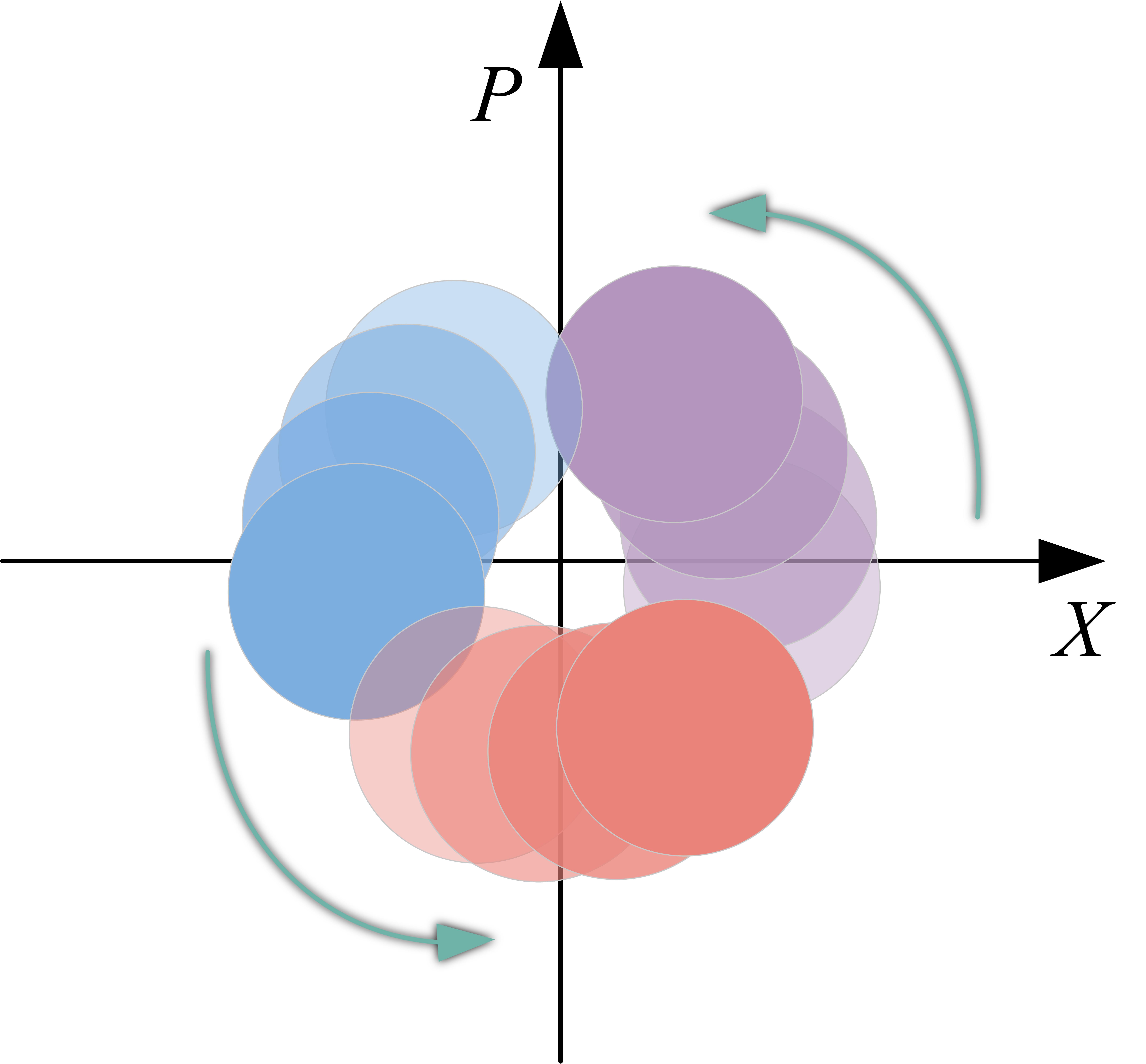}
		\label{fig1:b}
	}
	\caption{(a) Schematic diagram of the complementary modulation and measurement. The modulated quantum states have the same intensity $\mu$  but different phases.  In each  QRNG execution round, two successive temporal modes of coherent states are modulated with a $\pi$ phase difference. The early mode ($\rho_1, \rho_2$ or $\rho_3$ ) serves as the prepared state in the protocol, while the corresponding late mode ($\rho_1', \rho_2'$ or $\rho_3'$ )  acts as an auxiliary  state to balance the DC level of the homodyne detectors. The measured results of the two quadratures $(X,P)$ are discretized into $D_1$, $D_2$, $D_3$ and $D_4$. (b) Impact of phase drifts on observed states in phase space, where the prepared states undergo a random rotation over time.}
	\label{fig:1}
\end{figure}
\begin{figure*}[t]
	\centering
	\subfigure[]{
		\includegraphics[width=3.35in, clip]{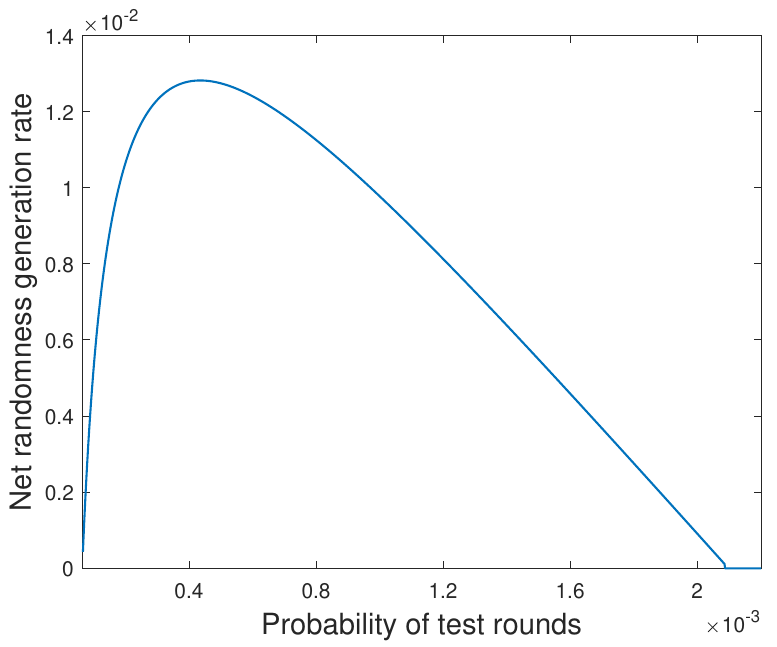} 
		\label{fig:a}
	}
	\hspace{0.000\textwidth} 
	\subfigure[]{
		\includegraphics[width=3.35in, clip]{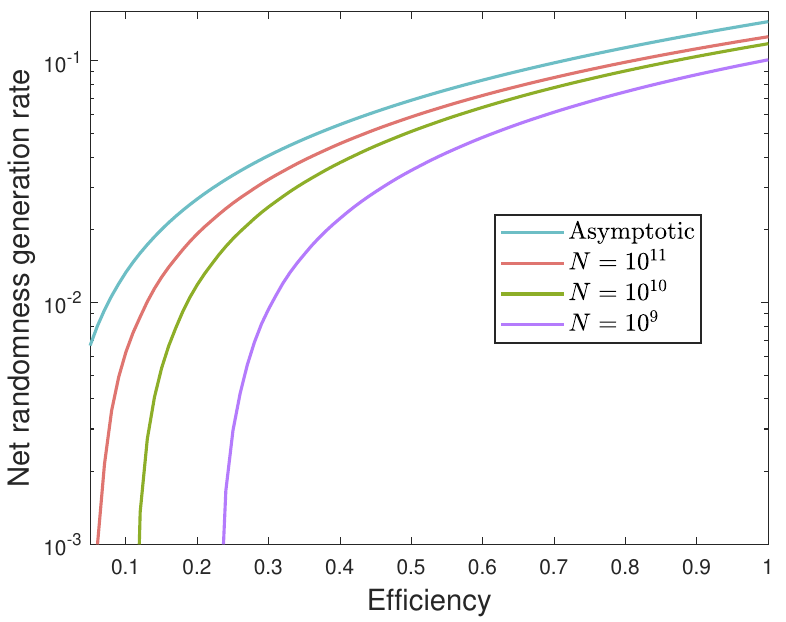}
		\label{fig:b}
	}
	\caption{(a) Simulations of the net randomness generation rate versus the probability of test rounds, with $\mu = 0.005$, $\eta=23.2\%$, $N=5.3\times10^9$ and $\epsilon = 10^{-10}$.  (b) Simulations of the net randomness generation rate versus detection efficiency for different  total number of rounds, with $\epsilon = 10^{-10}$, optimized mean photon number and  probability of test rounds.}
	\label{fig:double}
\end{figure*}
 
\subsection{\label{sec:citeref}Implementation
}
\hspace*{1em}To implement the proposed protocol, the source station randomly prepares coherent states $\lvert\sqrt{\mu}e^{ix\frac{2\pi}{3}}\rangle$ with the same probability ($p_{x} = \frac{1}{3}, x \in \{1,2,3\}$) in the test rounds, and  prepares a fixed state  $\lvert\sqrt{\mu}\rangle$  in the generation rounds. The prepared states are directed to heterodyne detectors in the measurement station, where the amplitude quadrature $X$ and the phase quadrature $P$ of the optical field  are measured. Depending on the region of the phase space in which  $(X,P)$ falls, the measured results $(X,P)$ are mapped into four disjoint sets $D_y$ defined by:
\begin{equation}
\begin{split}
D_y = \biggl\{(X,P) \;\biggm|&\; \sqrt{X^2 + P^2} \in (0,\infty), \\
&\tan^{-1}\left(\frac{P}{X}\right) \in \left[\frac{\pi}{2}(y-1), \frac{\pi}{2}y\right) \biggr\},
\end{split}
\label{eq:split_partition}
\end{equation}
where $y \in \{1,2,3,4\}$. The preparation and discretized measurement scheme of our protocol is illustrated in Fig.~\ref{fig:1}(a). An ideal heterodyne detection can be characterized by a POVM $\Pi_{\alpha} = \frac{1}{\pi} \lvert\alpha\rangle\langle\alpha\rvert$, where $\alpha = X + iP$ is the complex amplitude of the coherent state $\lvert\alpha\rangle$. For an input state \(|\sqrt{\mu} e^{ix\frac{2}{3}\pi}\rangle\), the probability density of the outputs $(X,P)$ of the heterodyne detection is given by the well-known Husimi function:
\begin{equation}
Q(\alpha)  = \frac{1}{\pi} e^{-\left|X+iP-\sqrt{\eta\mu} e^{ix\frac{2}{3}\pi}\right|^2},
\label{eq:husimi}
\end{equation}
where \(\eta\) is the total detection efficiency. In our case, $\eta$ can be expressed as $\eta=\eta_q\eta_e$, where $\eta_q$ is the quantum efficiency of the photodiodes and $\eta_e$ is  the equivalent efficiency induced by the electronic noise~\cite{appelElectronicNoiseOptical2007}.  Consequently, the theoretical conditional probability distribution \(p'(y|x)\) can be calculated as:  
\begin{equation}
\begin{split}
p'(y|x) &= \iint\limits_{D_y} Q(\alpha) \, dX \, dP \\
&= \frac{e^{-\eta\mu}}{\pi} \iint\limits_{D_y} e^{-r^2 + 2r\sqrt{\eta\mu}\cos\left(\varphi - x\frac{2}{3}\pi\right)} r\,dr\,d\varphi.
\end{split}
\label{eq:conditional_prob}
\end{equation}
In cases of collective attacks, we can substitute \(p'(y|x)\) for \(p(y|x)\) in Eq.~(\ref{eq:constraints}) and obtain the expected performance of the measurement device by solving the SDP. In cases of general attacks, \(p'(y|x)\) can be substituted as the nominal value in Eq.~(\ref{eq:dual_problem}) to obtain the solution $\{\omega_{xy}'\}$ of the dual SDP. It is worth noting that $\omega_{xy}'$ is closely related to the randomness bound $R_{\mathrm{net}}$.  Herein, we introduce an constraint on $\omega_{xy}'$, namely $\omega_{xy}' \leq 0$, which still satisfies Eq.~(\ref{eq:solution_set}) and therefore preserving security. Upon completion of the above constraints, we can apply Kato's inequality to the constructed random variables $W_n^{\text{nor}}$, with the parameters $a$ and $b$ from Eq.~(\ref{eq:kato-in})  optimized to tighten $R_{\mathrm{net}}$, subject to the constraints
\begin{equation}
\begin{aligned}
\exp\left( -\frac{2(b^2 - a^2)}{\left(1 - \frac{4a}{3\sqrt{N}}\right)^2} \right) &= \epsilon, \\
\sqrt{N} - 2a > 0, \quad a \in \mathbb{R}, \quad b &\geq 0.
\end{aligned}
\label{eq:const-a-b}
\end{equation}

Based on the  definition of net randomness generation rate $R_{\mathrm{net}}$  in Eq.~(\ref{eq:net-rate}), we simulate the effects of key parameters on the performance of our protocol, where we set $N_{y|x} = p_xp_tNp'(y|x)$ and the violation probability of Kato’s inequality $\epsilon = 10^{-10}$. Fig.~\ref{fig:double}(a) shows the $R_{\mathrm{net}}$   as a function of the  probability of  test rounds. Fig.~\ref{fig:hmin} shows the simulations of $R_{\mathrm{net}}$ and  the associated randomness cost versus  mean photon number. Fig.~\ref{fig:double}(b) plots the relation between $R_{\mathrm{net}}$  and $\eta$  for different  total number of rounds $N$, where the mean photon number and  probability of test rounds  are chosen optimally. As $N$ increases, the $R_{\mathrm{net}}$ converges to the asymptotic limit $H_{\min}$. This is because when $N \to +\infty$, the correction term in Eq.~\eqref{eq:delta-f} approaches zero, hence $H'_{\min}$ in Eq.~\eqref{eq:min_entropy_fin} reduces to  $H_{\min}$ in Eq.~\eqref{eq:min_entropy}, yielding $R_{\mathrm{net}}\approx H'_{\min}\approx H_{\min}$. Here, $p_t$ can be chosen small enough such that $R_{\mathrm{in}}$ is negligible. However, real-time randomness extraction in real-world applications necessitates finite block lengths to sustain throughput. In this case, both finite-size and non-i.i.d. effects become significant and non-negligible.
\begin{figure*}[t]
	\centering
	\includegraphics[width=1\linewidth]{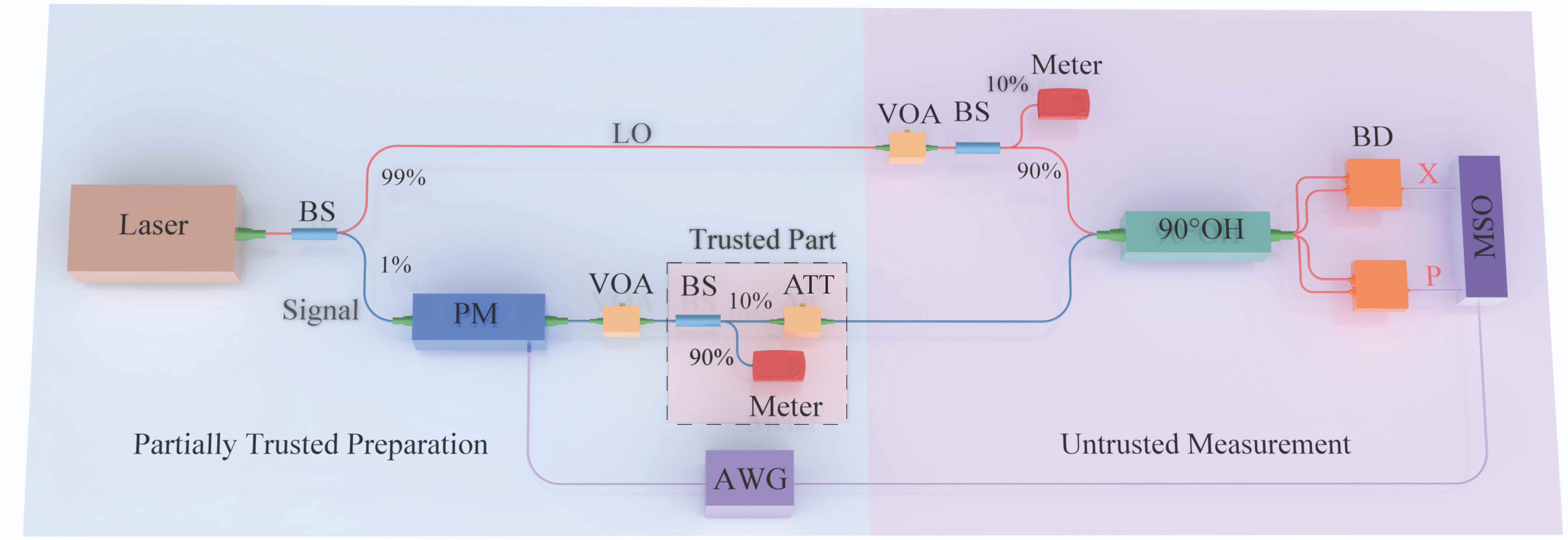}
	\caption{Experimental setup of the Semi-DI QRNG. The phase of the input signal is randomly modulated by a PM, while its amplitude is evaluated through the Meter reading from the trusted part. After optimizing the intensity of LO, the prepared state is interfered with the LO in a 90° OH, followed by measurement of two quadratures using a pair of BDs. Through the statistical dependence between the prepared states and measurement outcomes, the lower bound on the generated randomness is established. BS: beam splitter; VOA: variable optical attenuator; Meter: optical power meter; PM: phase modulator; ATT: fixed optical attenuator; 90° OH: 90° optical hybrid; BD: balanced amplified photodetector; AWG: arbitrary waveform generator; MSO: mixed-signal oscilloscope.}
	\label{fig:enter-label}
\end{figure*}

As pointed out in Sec.~\ref{sec:level2}, a truly random input is optimal for the execution of our QRNG protocol.  Given this, to show the practical utility of our protocol, we need to prove that it  generates more randomness than it consumes. In our work, the input  randomness serves three tasks: state preparation in test rounds, switch between generation and test rounds and the seed for the randomness extractor. As we employ a strong extractor whose seed is considered part of its output, the randomness consumption is thus attributed  to the first two components. Therefore, following the analysis in Ref.~\cite{wangProvablysecureQuantumRandomness2023a}, one could employ the interval algorithm \cite{Intervalalgorithm} to convert a uniform bit string into the required biased bit string. In this scenario, the average input randomness per round $R_{\mathrm{in}}$  of our protocol  (for large $N$) is given by 
	\begin{equation}
		R_{\text{in}} = h_2(p_t) + p_t H(x) ,
		\label{eq:R_in}
	\end{equation}
	where $h_2(\cdot)$ is the binary entropy function and $H(x)$ is the Shannon entropy of the input distribution $\{ p_x  \}$.

\section{EXPERIMENT}
\hspace*{1em}The proposed protocol is demonstrated with a CV fiber system. The experimental setup is illustrated in Fig.~\ref{fig:enter-label}. The system consists of two main parts: partially trusted quantum state preparation and untrusted quantum state measurement. All optical components are connected via polarization-maintaining fibers and commercially available.

In the quantum state preparation, a laser module (Laser) emits continuous-wave light with wavelength of 1538 nm, which is split into two paths by a 99:1 beam splitter (BS). The major portion serves as the LO for heterodyne detection, and the minor portion is the signal for the state preparation. In the signal path, the light is injected into a LiNbO$_3$ phase modulator (PM) with a bandwidth of 300 MHz. An arbitrary waveform generator (AWG) operating at 100 MHz randomly prepares coherent states with phases of $0$, $\frac{2}{3}\pi$ and $\frac{4}{3}\pi$. The modulated states are then attenuated using a variable optical attenuator (VOA) and split into two paths by a 90:10 BS. 90\% of the light is directed into an optical power meter (Meter, Thorlabs S132C with a resolution of 1 nW), while the remaining 10\% is connected to a fixed attenuator (ATT). The prepared states $\lvert\sqrt{\mu}e^{ix\frac{2\pi}{3}}\rangle$ are directed to the measurement station. By combining the power meter readout and the precalibrated attenuation factor of the ATT, the intensity of the prepared optical field is characterized. Here, we use a time window of 10 ns (time duration per quantum state) to estimate the mean photon number $\mu$ of prepared states. The monitoring the energy of quantum states is considered trusted.

\begin{figure}
	\centering
		\includegraphics[width=1\linewidth, height=2.5cm]{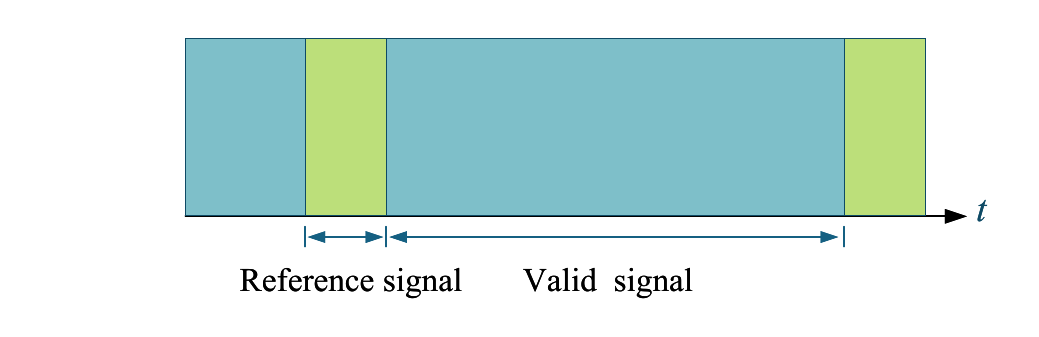}
	\caption{Data frame configuration of QRNG experiment. Reference signals for data synchronization are periodically inserted between the valid signals used for QRNG execution. The valid signals consist of the prepared and auxiliary signals required for the complementary modulation.}
	\label{fig:complete_figure}
\end{figure}

In quantum state measurement, the LO passes through a VOA that adjusts the light intensity, and is then split into two beams by a 90:10 BS. 10\% of the light is directed to the Meter for the purpose of monitoring the power of LO. Notice that the protocol requires no precise characterization of the measurement, therefore, the LO is monitored not for security purposes but merely for enhancing the performance of the QRNG in this proof-of-principle experiment. The remaining 90\% of the LO and the prepared states are injected into the LO port and the signal port of the 90° optical hybrid (90° OH), respectively. Two balanced amplified photodetectors (BDs), each with a bandwidth of 1.6 GHz, measure the two quadratures of the prepared states. By careful calibration, the efficiency  of the heterodyne detection in our experiment is found to be 23.2\%. The output signals from the BDs are sampled by a Tektronix MSO5204B oscilloscope (MSO) with sampling frequency of 1 GHz. To realize a synchronization sampling, the MSO and AWG share a reference clock and the MSO is triggered by the AWG. Finally, the acquired data is sent to a computer for offline postprocessing. To reconstruct the signal states, ten samples are averaged to represent a single signal state.

The parameters employed in the experiment  are presented in Table~\ref{tab:exp_params}, where the mean photon number of the prepared states $\mu$ and the probability of test rounds $p_t$ have been optimized. The proportion of generation rounds is chosen to be large to increase the randomness generation rate. This implies that the modulator spends the majority of its time preparing a fixed state, which induces a DC imbalance in two AC‑coupled  homodyne detectors and causes  distortions of quantum signals ~\cite{MacDermott1998}. To mitigate this effect, we adopt the complementary modulation scheme following the approach in Ref.~\cite{wangProvablysecureQuantumRandomness2023a}, which is presented in Fig.~\ref{fig:1}(a). In our scheme, each prepared state is paired with an auxiliary state exhibiting a $\pi$ phase difference within each round. Thus,  the two temporal modes (the prepared  and auxiliary state)  yield opposite expectation values for the same  quadrature measurements, thereby balance the DC output of two homodyne detectors.  In addition, to precisely locate the valid signal segments  used for QRNG execution, we insert a predefined reference signal before each valid signal segment. The data synchronization process is implemented by searching the maximum  correlation between the reference signal and the detected signal. The  data frame configuration of our experimental implementation is shown in Fig.~\ref{fig:complete_figure}.

\begin{table}[t]
	\centering
	\caption{Experimental parameters}
	\begin{tabular}{ccc}
		\doubletoprule
		Mean photon number $\mu$   & 0.005\\
		\addlinespace[2pt]
		Probability of test rounds $p_t$ &$3.464\times10^{-4}$  \\
		\addlinespace[2pt]
		Total number of rounds $N$& $5.3\times10^9$\\
		\addlinespace[2pt]
		Detection efficiency $\eta$  & 23.2\%\\
		\addlinespace[2pt]
		Input distribution in test rounds $p_x$& $\frac{1}{3}, \forall x$\\
		\addlinespace[2pt]
		Violation probability of Kato’s inequality $\epsilon$ &$ 10^{-10}$  \\
		\doublebottomrule
	\end{tabular}
	\label{tab:exp_params}
\end{table}
\vspace{-2pt}

\begin{figure}[]
	\centering
	\includegraphics[width=1\linewidth]{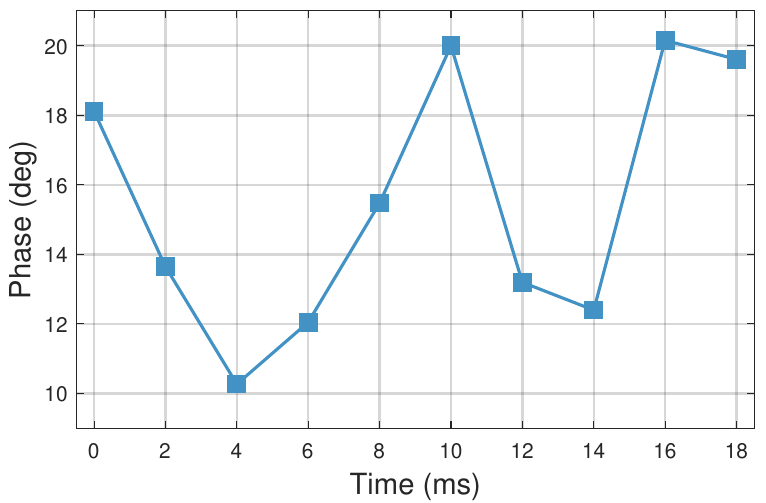}
	\caption{Phase drift of the prepared state $\rho_3$  versus time. The phase  is evaluated at intervals of $2\,\mathrm{ms}$.}
	\label{fig:phase}
\end{figure}

In practical implementations, the relative phase $\theta$ between the quantum signals and local oscillator (LO) may drift over time due to environmental variations, or modulator instability. Consequently, the phases of the three prepared coherent states deviate from their ideal values of $\frac{2}{3}\pi$, $\frac{4}{3}\pi$ and $2\pi$, and the observed quantum states rotate in the phase space, as illustrated in Fig.~\ref{fig:1}(b). This leads to a reduction in the extractable randomness. Because such a phase drift can be intuitively considered equivalent to classical noise in $N_{y|x}$ in Eqs.~\eqref{eq:constraints} and~\eqref{eq:Ng}. Considering that the time scale of the phase drift is much slow than that of the sampling rate, therefore, the sampled data sequences can be divided into multiple blocks. For each block that contains three quantum signals, the phases of the quantum signals remain nearly constant. To optimize phase-tracking accuracy, we select a block of 10000 prepared states based on the experimental observation that the duration of each block (200 $\mu\mathrm{s}$) is slightly shorter than the characteristic time of the phase fluctuations (the order of millisecond-scale), as shown in Fig.~\ref{fig:phase}. By averaging $(X, P)$ of $\rho_3$ (which is prepared in most of the time) in each block, the shot noise can be suppressed and the phase of $\rho_3$ can be determined by $\theta = \tan^{-1} \frac{\overline{P}}{\overline{X}}$.  Therefore, the three quantum signals  within each block are rotated by $-\theta$ to correct the phases to the configuration shown in Fig.~\ref{fig:1}(a). This allows us to assign a measured output $y$ to each $(X,P)$ group.

\section{RESULTS}
\begin{figure}[]
	\centering
	\includegraphics[width=1\linewidth]{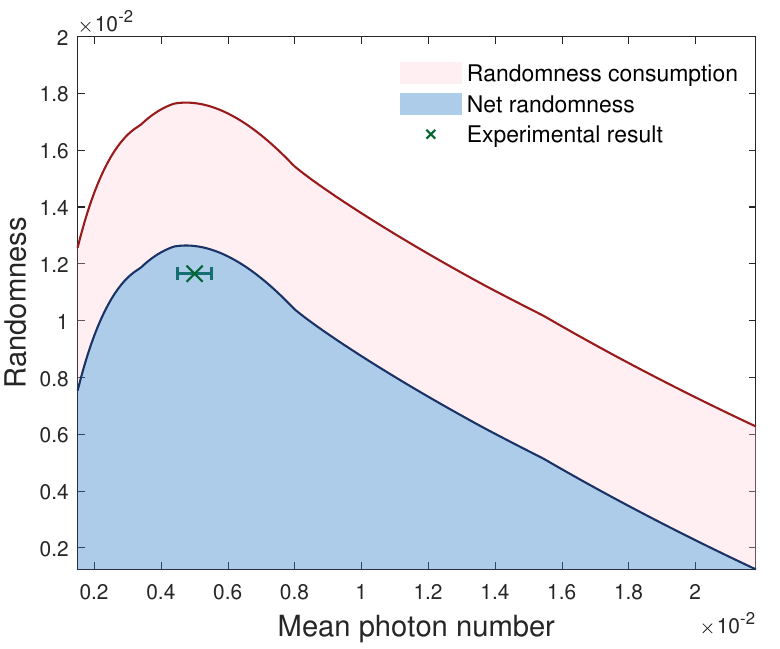}
	\caption{Randomness versus mean photon number with $\eta=23.2\%$,  $p_t=3.464\times10^{-4}$, $N=5.3\times10^9$ and $\epsilon = 10^{-10}$. Red and blue curves denote the expected gross and net randomness generation rates, respectively. The pink area shows the associated randomness consumption. Green cross represents  our experimental result, where the error bar denotes the estimation precision on mean photon number.}
	\label{fig:hmin}
\end{figure}

\hspace*{1em}In our experiment, with the mean photon number $\mu=0.005$, we run  a total of $5.3\times10^9$ rounds,  of which about $1.8\times10^6$  rounds are selected as test rounds to obtain the statistics $N_{y|x}$. Hence, based on Eqs.~(\ref{eq:Ng}) and (\ref{eq:min_entropy_fin}), the gross randomness generation rate $R_{\mathrm{gross}}$ obtained by our QRNG execution is estimated to be  0.01668 bits per round. Considering the consumed randomness for choosing test rounds and prepared states, the average randomness consumption rate $R_{\mathrm{in}}$ in our case  evaluated via Eq.~(\ref{eq:R_in}) is 0.00503 bits per round. Consequently, our QRNG system achieves a net randomness generation rate of 0.01165 bits per round (green cross in Fig.~\ref{fig:hmin}).

Employing a two-bit encoding scheme, we convert the sequence $y$ into a binary sequence (raw bit string). A secure random number sequence is then generated from the raw bit string according to $H'_{\min}$ with a strong randomness extractor based on a Toeplitz hash function~\cite{tomamichelLeftoverHashingQuantum2011,Krawczyk1994,Krawczyk1995}. Finally, we perform the randomness tests on the generated bit sequence by using the NIST Statistical Test Suite and find that the random sequence passes the NIST tests.
\vspace{\baselineskip}
\section{CONCLUSION}
\hspace*{1em}In this study, we implement a semi-DI QRNG based on relaxed assumptions with commercially available components. Our protocol requires an easily verifiable energy-bound assumption on the source and allows the adversary to perform non-i.i.d. general attacks at the measurement side. Finite-size effects are also incorporated into the security analysis.  Compared to the randomness estimation presented in Refs.~\cite{braskMegahertzRateSemiDeviceIndependentQuantum2017b,tebyanianSemideviceIndependentRandomness2021a,avesaniSemiDeviceIndependentHeterodyneBasedQuantum2021a}, our work explicitly takes into account the randomness overhead and demonstrates a positive net randomness generation rate, thereby enhancing its practical feasibility. In addition, heterodyne detection is employed, enabling phase compensation through postprocessing without the need for an active phase stabilization system, which reduces the complexity of QRNG implementation.

Operating at 100 MHz, our QRNG system achieves a secure net random number generation rate of 1.165 Mbps, offering a promising approach for  high security, robust and high generation rate semi-DI QRNG.

In future work, we will improve the system repetition rate to increase the generation rate of the QRNG. It is expected that the protocol can be extended to more input states and discretization outputs in CV system. In this case, the minimum entropy per round can be further increased.
\vspace{\baselineskip}
\section*{ACKNOWLEDGMENTS}
\hspace*{1em}We thank  S. Liu and H. Shi for helpful discussions. This work was supported by National Natural Science Foundation of China (Grants No. 62205188 and No. 62175138) and the Quantum Science and Technology-National Science and Technology Major Project (Grant No. 2021ZD0300703).

% The \nocite command causes all entries in a bibliography to be printed out
% whether or not they are actually referenced in the text. This is appropriate
% for the sample file to show the different styles of references, but authors
% most likely will not want to use it.
%

%\bibliography{ref-semi1}
\end{document}